\documentclass[a4paper,10pt]{article} 
\usepackage{jcappub}

\usepackage[english]{babel}
\usepackage{setspace}
\usepackage{graphicx}
\usepackage{subfig}
\usepackage{booktabs}
\usepackage{tabularx}
\usepackage{caption}
\usepackage[latin1]{inputenc}
\usepackage{color}

\newcommand{\ie}{{\it i.e.}~}
\newcommand{\eg}{{\it e.g.}~}

\newcommand{\lcdm}{$\Lambda$CDM\,}
\newcommand{\ar}{\alpha_{\mathrm{Ric}}}
\newcommand{\as}{\alpha_{\mathrm{Scal}}}

\newcommand{\F}{F(X,\varphi)}
\newcommand{\lint}{\mathcal{L}^{nmc}}

\newcommand{\ros}{\rho_*}
\newcommand{\fai}{{\bf\Phi}}
\newcommand{\psai}{{\bf \Psi}}
\newcommand{\Hc}{\mathcal{H}}

\title{Non-minimally coupled dark matter: effective pressure and structure formation}

\author[a,b]{Dario Bettoni,}
\author[a,c]{Valeria Pettorino,}
\author[a,b]{Stefano Liberati}
\author[a,b]{ and Carlo Baccigalupi}
\affiliation[a]{SISSA/ISAS,
Via Bonomea 265, 34136, Trieste, Italy}
\affiliation[b]{INFN, Sezione di Trieste, 
Via Valerio, 2, 34127, Trieste, Italy}
\affiliation[c]{D\'epartement de Physique Th\'eorique, Universit\'e de
Gen\`eve, \\
Quai E. Ansermet 24, CH-1211 Gen\`eve 4, Switzerland}
\emailAdd{bettoni@sissa.it}
\emailAdd{Valeria.Pettorino@unige.ch}
\emailAdd{liberati@sissa.it}
\emailAdd{bacci@sissa.it}

\abstract{
We propose a phenomenological model in which a non-minimal 
coupling between gravity and dark matter is present in order to address some of the apparent small scales issues of \lcdm model. When described in a frame in which gravity dynamics is given by the standard Einstein--Hilbert action, the non--minimal coupling translates into
an effective pressure for the dark matter component.
We consider some phenomenological examples and describe both background and linear perturbations. We show that the presence of an effective pressure may lead these scenarios to differ from \lcdm at the scales where the non--minimal coupling (and therefore the pressure) is active. In particular two effects are present: a pressure term for the dark matter component that is able to reduce the growth of structures at galactic scales, possibly reconciling simulations and observations; an effective interaction term between dark matter and baryons that could explain observed correlations between the two components of the cosmic fluid within Tully--Fisher analysis.
}


\date{\today}

\begin{document}

\maketitle 
\flushbottom

\section{Introduction}

Our present understanding of the universe is encoded in the so called \lcdm model which is able to explain in a simple and elegant way the evolution and the formation of structures.
However, despite its great successes \cite{WMAP} and successful tests at the solar system scale \cite{GR}, General Relativity is still far from being the ultimate theory of our universe. In particular, the present day energy density is dominated by dark fluid(s) whose nature has not been completely understood so far.
At the largest scales the accelerated expansion is still a puzzle to be solved while, at smaller scales, galaxy clusters and galaxies are dominated  by a dark matter component whose particle nature is still unknown.

Dark matter, introduced in the realm of astrophysics in 1934 by Fritz Zwicky \cite{Zwicky}, is still missing both a definitive detection and a unique theoretical candidate. Over the years many particles have been proposed as viable candidates for dark matter (see \cite{Bertone,Feng, Bergstrom} for comprehensive presentations) but up to now no convincing indirect \cite{Cirelli, Pamela,Fermi} nor direct \cite{Gaitskell,DAMA,CRESST,COGENT,CMS} detection has led to a resolutive answer.
At cosmological and astrophysical scales there is no need for a particle description of dark matter and its dynamics is described via a fluid approximation. This fluid is obtained from the particle description through some averaging procedure that keeps track only of those microscopic properties that are relevant from a macroscopic point of view, like pressure, velocity and density.

The most accredited model, the Cold Dark Matter paradigm (CDM), describes dark matter as a non--interacting, pressureless fluid with small dispersion velocity. As already pointed out, despite its simplicity this model is able to describe the formation and evolution of structures with remarkably high precision.
However, although its success, recent observations seem to indicate that the CDM picture may not be able, as it stands, to explain in a natural way dark matter dynamics at clusters and galactic scales.
In particular some tension exists between dark matter simulations and observations, with regard to both the density profiles of dark matter halos \cite{cuco,NFW,Li,Simon,Zavala} and for the number of predicted substructures inside a given host halo \cite{Bullock,Strigari,Klypin,Navarro}.\footnote{The possibility that these issues may be related to a too low resolution of the simulations has been investigated too \cite{Mayer}.} Indeed, many attempts have been made to solve these issues: baryons feedback on dark matter distribution \cite{AGN,SN,Dyneff,Gov}, self interacting dark matter \cite{Spergel, bullett,bullett2}, fuzzy cold dark matter \cite{Hu}, frustrated cold dark matter \cite{Varun} in the context of unified dark matter dark energy models or warm dark matter (WDM)  \cite{Bode,Boya,Viel1,Viel2, Maccio1,Maccio2}, whose free streaming length is able to erase structure formation below it.  These problems may be related to the lack of a minimum length scale for the clustering of dark matter that hence keeps collapsing down to small scales. In other words, the net effect can be interpreted as the insurgence of an effective pressure term in the dark matter fluid.  Recently it has been proposed a method \cite{Visser} to test the DM equation of state at small scales in a rather model independent way by comparing the gravitational potentials measured from galaxy rotational curves and from weak lensing which has been applied to data in \cite{Serra}. The results show a value for the equation of state parameter compatible with zero but with a preferred value of $w=-1/3$ at $1\sigma$. From cosmological observations, and hence large scales, the equation of state for DM is compatible with zero \cite{Muller,Calabrese}. Interestingly, even if  no incompatibility with w=0 has been found, a model with $w \neq 0$ is still compatible with CMB data \cite{Calabrese}.

Another source of tension concerns empirical tests like the baryonic Tully-Fisher relation \cite{TFR,BTFR} or the constant galactic surface density \cite{Salucci}: they both hint towards a connection between dark matter and baryonic properties; these observations are hardly explained within the \lcdm paradigm unless the amounts of dark matter and baryons are accurately balanced. On the other hand, dark matter--baryon interactions are strongly constrained in the context of particle physics so that it is very unlikely they can address these issues.
 
Curiosly, a model capable of predicting at least some of these empirical tests is the well known MOND paradigm \cite{MOND,MOND2} where the mass discrepancy problem is explained via a modified Newtonian dynamics without requiring dark matter. MOND is a non relativistic model and several generalizations to a fully relativistic theory have been attempted \cite{STVGT, Bim,bekitvsg,bekirelmond}. At present, none of them seems however able to reproduce with accuracy and most of all, simultaneously, all well established cosmological features, like the CMB acoustic 
peaks, the matter power spectrum, the Bullet cluster (see for example \cite{MONDCMB,skordis} for attempts in this direction).

An alternative appealing solution to CDM problems that may reconcile \lcdm successes at large scales with MONDian behavior at galactic scales, comes from a new class of non-standard dark matter gravity couplings \cite{RMDM,bruneton,Bettoni}.
In this scenario the fluid content of the \lcdm paradigm is maintained but  it is rather the dynamics of the dark matter fluid at suitable late times and small scales to be changed.
In particular, we shall here entail the idea that dark matter could, at suitable scales, have undergone some sort of phase transition (e.g. something analogue to a Bose-Einstein condensation) and consequently developed a finite coherence length of size comparable to the local spacetime curvature. In this sense dark matter would be capable to probe gravity - \eg to ``feel" second derivatives of the metric - hence becoming  non--minimally coupled to curvature terms. 

In this paper we further investigate the model proposed in \cite{Bettoni}  deriving for the first time its cosmological implications. We show how this class of models gives rise to two effects: a pressure term for dark matter, able to reduce the growth of structures at small scales, plus an effective interaction term between dark matter and baryons that might be able to explain  observed correlations between the two components.

The paper is organized as follows. In section \ref{sec:NMC} we introduce the model, in section \ref{sec:PDM} we describe the field/fluid formalism used and write the equations in their general form. In section \ref{sec:cosmology} we specialize them to a flat FRLW universe and study the modifications of our model with respect to \lcdm while in section \ref{sec:results} numerical results are presented. Finally in section \ref{sec:conclusions} we draw out our conclusions.

\section{Non--minimally coupled dark matter}
\label{sec:NMC}

The most general--scalar tensor theory that gives second order field equations is encoded in the Horndeski action \cite{Horndeski}. Unfortunately due to its generality, it is very hard to build a constrained and reliable cosmology using this action (see however \cite{defelice1,defelice2} for attempts in this sense). In \cite{Bettoni} a minimal recipe was used to build an action in which the dark matter fluid is non--minimally coupled to curvature terms and the astrophysical consequences of this couplings were addressed. The action reads:
\begin{multline}
S = \frac{c^{3}}{16 \pi G_N}\int d^{4}x\,\sqrt{-\tilde{g}}  (1+\alpha_{Scal}\Upsilon(\rho_{DM}))\tilde{R}+\\
 \frac{\ar c^{3}}{16 \pi G_N}\int d^{4}x\,\sqrt{-\tilde{g}} \tilde{R}_{\mu \nu}\Xi(\rho_{DM})u_{DM}^{\mu}u_{DM}^{\nu} + S_{DM}[\tilde{g},\rho_{DM}]+S_{B}[\tilde{g},\rho_B].
\label{eq:JFaction}
\end{multline}
 where $\tilde{g}$ is the metric in the Jordan frame (we will use the tilde to indicate quantities in the Jordan frame), c is the velocity of light, $G_N$ is the Newton constant, $\tilde{R}$ and $\tilde{R}_{\mu \nu}$ are the curvature scalar and Ricci tensor respectively in the metric $\tilde{g}$, $\as$ and $\ar$ are constants, $\rho_{DM}$ is the Dark Matter (DM) energy density, $\rho_B$ is the baryonic energy density. 
 $\Upsilon(\rho_{DM})$ and $\Xi(\rho_{DM})$ are two generic functions of the DM density.  Here $\as$ introduces a coupling between $\rho_{DM}$ and the curvature; in addition, $\ar$ couples the metric and its derivatives also to the DM four velocity $u_{DM}$. 
This model gives rather precise predictions in the Newtonian limit: the Newtonian gravitational potential $\tilde{\psai}$ is sourced not only by the energy density but also by its gradient and Laplacian:
\begin{equation}
\nabla^{2} \tilde{\psai} = 4 \pi G_{N} \left( \rho_{DM} -\frac{ \ar}{2}  \nabla^2 \Xi(\rho_{DM}) + \as \nabla^2 \Upsilon(\rho_{DM}) \right).
\end{equation}
 These features are interesting ingredients in the tentative solution of \lcdm small scales issues as they generate corrections that may reduce the cuspiness of the DM density profiles and reduce the number of satellite halos.
We note that our model was developed in a context which is independent from Horndeski; it can still be seen as a subclass of those theories, when only fluid quantities related to density and quadrivelocity and their coupling to curvature terms are taken into account.

We will now further extend the analysis of this class of models, investigating their cosmological consequences. In order to do this we choose to move to a frame in which the action for gravity is described by the standard Hilbert--Einstein action. When in the Jordan frame the non--minimal coupling is taken to be with the Einstein tensor, $ \as = 1/2 \ar \equiv \epsilon$, this can be achieved with the following metric transformation:
\begin{equation}
\tilde{g}_{\mu \nu} = g_{\mu \nu} + h_{\mu \nu}, \qquad h_{\mu \nu} = \left(A(\rho_{DM})^2-1\right)g_{\mu \nu} + B(\rho_{DM}) u_{\mu}u_{\nu},\label{disfrel}
\end{equation}
where $A(\rho_{DM})$ and $B(\rho_{DM})$ are generic functions of the DM density; this is the most general relation between metrics that respect both causality and the equivalence principle (WEP) and is called disformal transformation \cite{disf}. We stress that here $h_{\mu \nu}$ is not only a function of the metric but depends also on the DM density itself. If we re-express action (\ref{eq:JFaction}) in terms of the metric $g_{\mu \nu}$ and then expand in powers of $h_{\mu \nu}$ up to order $\mathcal{O}(h_{\mu \nu}^2)$ we obtain the following action:
\begin{equation}
S=S_{HE}[g]+S_{DM}[g,\rho_{DM}] + S_B[g,\rho_B] +S_{int}[g,\rho_B,\rho_{DM}],
\label{action}
\end{equation}
where $S_{HE}[g]$ is the standard Einstein--Hilbert gravitational action, $S_{DM}[g,\rho_{DM}]$ and $S_{DM}[g,\rho_{DM}]$ are the DM and baryonic actions in the metric g, $S_{int}$ is
\begin{equation}
S_{Int}=- \frac{1}{2}\int d^4x\sqrt{-g}\left(T^{\mu \nu}_{DM}+T^{\mu \nu}_{b}\right)h_{\mu \nu}(g,\rho_{DM}),
\end{equation}
and represents a new interaction term which in general involves the metric, dark matter and baryons. As usual the stress energy tensor $T_{\mu \nu}$ for the i component (DM or baryons) is given by
\begin{equation}
T_{\mu \nu}^i = -\frac{2}{\sqrt{-g}}\frac{\delta S_{i}}{\delta g_{\mu \nu}}.
\end{equation}
In this new frame the effects of the non--minimal coupling have been transferred into a coupling term for the stress energy tensors of both dark matter and baryons and $h_{\mu \nu}$ which, as stressed before, is itself a function of the metric as well as of dark matter fluid variables. This translates directly into a coupling between DM and baryons and a self coupling for dark matter. Notice that those two couplings are not to be intended in the particle physics sense but they rather emerge from a geometrical coupling between dark matter and gravity. 

We further stress here that  the action (\ref{action}) is obtained from (\ref{eq:JFaction}) through an expansion and hence the two actions are equivalent up to order $\mathcal{O}(h)$. This doesn't matter as we could have started directly from the action (\ref{action}).

As argued in the introduction we assume
the coupling to be active at late times and small scales when dark matter dynamics changes giving rise to a coherence length that activates the non--minimal coupling.

\section{Dark matter with an effective pressure}
\label{sec:PDM}

As a first investigation of this model we neglect the baryon 
contribution and set the value of the conformal function in eq. (\ref{disfrel}) to be $A=1$. 
We also switch to field formalism using standard conventions  \cite{Weinberg}:
\begin{equation}
\rho \rightarrow \rho(X,\varphi) , \qquad p \rightarrow p(X,\varphi), \qquad u_\mu=\nabla_\mu \varphi/\sqrt{2X}
\end{equation}
where now $\varphi$ is the dark matter scalar field and $X=- g^{\mu \nu}\nabla_\mu\varphi \nabla_\nu\varphi/2$ is its kinetic part. 
The field variables are not fundamental fields, but rather stand 
for a different representation of the fluid. In this sense we must ensure that in absence of a coupling we recover the equations for the standard dark matter fluid. In general a scalar field will not behave as a pressureless dust; 
thus, we have to enforce this requirement. In order to do this we introduce a Lagrangian multiplier in the Lagrangian for the dark matter field
\begin{equation}
\mathcal{L}=\lambda(x^\mu)\left(X-\frac{1}{2}V^2(\varphi)\right)
\end{equation}
where $V(\varphi)$ is a generic potential so that when we take the variation of the action with respect to $\lambda$ we get
\begin{equation}
\frac{\delta S_{DM}}{\delta \lambda}=0 \Rightarrow X-\frac{1}{2}V^2(\varphi)=0
\end{equation}
which exactly sets the pressure to zero \cite{Iggy,lagmult}.
We conjecture that in order for the interaction to be active, the temperature of the cosmological bath (\ie time or redshift) must be below a certain threshold. This explains the time dependence of the coupling. However, even after the critical temperature has been reached, the density may be too low for the interaction to be efficient. Hence one needs high densities in order to make the interaction relevant.
In order to implement phenomenologically the wished scale and time dependences of the coupling, we fix the parametrization of the coupling function in the following way:
\begin{equation}
h_{\mu \nu}(X,\varphi)= \frac{F(X,\varphi)}{\ros}\nabla_\mu\varphi\nabla_\nu\varphi.
\end{equation}
The function $F$ gives the scale dependence of the interaction by suppressing it until the density is large enough to overcome a given threshold fixed by $\ros$.

The action (\ref{action}) can then be rewritten as:
\begin{equation}
S=S_{HE}[g]+S_{DM}[g,\varphi,\lambda] + \epsilon\int d^4x\sqrt{-g}\lint(X,\varphi)
\label{mainaction},
\end{equation}
where now $S_{DM}$ depends on $\{\varphi,\lambda\}$ rather than on $\rho_{DM}$ and
\begin{equation}
\lint= -\frac{F(X,\varphi)}{\ros}X\left(X+\frac{1}{2}V^2(\varphi)\right)
\end{equation}
is the interaction term generated by the non--minimal coupling.
The coupling $\epsilon$ is a switch that is chosen to be zero at $t<t_c$, $t_c$ being the time when the coupling is activated; for $t>t_c$, $\epsilon$ becomes different from zero, reaching a constant value of about $\epsilon<1$, as required by the expansion of $h_{\mu \nu}$ to order $(\epsilon^2)$. In other words, $\epsilon$ becoming different from zero indicates the onset of the non--minimally coupled epoch. At the same time, for the purpose of the present analysis, which is limited to order $\mathcal{O}(\epsilon^2)$, its value must be small compared to one.\footnote{The parameter $\epsilon$ is just a phenomenological parameter that should be given dynamically by the non--minimal coupling mechanism. In
this sense we are not introducing a non--dynamical field and background independence is preserved.} From a cosmological point of view the time dependence of the coupling is required because we want to study how the non--minimal coupling modifies the \lcdm behavior at small scales and late times only, as we stressed in the introduction. To be fully rigorous we should have given $\epsilon$ a spatial dependence as well. That would cause the activation of the non--minimal coupling to happen at different times and in different regions. Such a spatial dependence would give rise to a most interesting phenomenology but strongly dependent on which powering mechanism is chosen for such a dependence on space. We therefore feel that this treatment would go beyond the scope of the present paper, requiring a dedicated analysis in a separated work.

The variation with respect to $\varphi$ and $\lambda$ gives the two combined equations of motion which together specify the fluid dynamics:
\begin{eqnarray}
\dot{\lambda} & = & V^{-2}(\varphi)\left[V(\varphi)\rho_{,\varphi}-(\rho + \epsilon p)\theta\right],\label{llfluid}\\
\dot{\varphi} & = & -V(\varphi),
\label{lffluid}
 \end{eqnarray}
 where
 \begin{equation}
 \dot{()}\equiv u_\mu \nabla^\mu, \qquad \theta \equiv \nabla_\mu u^\mu,
 \end{equation}
and where $\rho_{,\varphi}$ is the derivative of the density with respect to the field $\varphi$. The energy density $\rho$ and  pressure $p$ are derived from the total dark matter stress energy tensor:
\begin{equation}
T_{\mu \nu} = \left(\lambda+\epsilon \lint_{,X}\right)V^2(\varphi)u_\mu u_\nu - \epsilon\lint g_{\mu \nu}
\end{equation}
by a direct confrontation with the form ad the perfect fluid stress energy tensor. We can identify the following thermodynamic quantities:
\begin{eqnarray}
\rho & = & \left(\lambda +\epsilon\lint_{,X}\right)V^2(\varphi)-\epsilon\lint,\label{ro}\\
p & = & \lint \label{pressure} ,\\
u^\mu & = & V^{-1}(\varphi)\nabla^\mu \varphi .
\label{thermo}
\end{eqnarray}
Notice that these quantities are not the same appearing in action (\ref{action}) because here the interaction term is directly involved in the definition of both density and pressure. In other terms in (\ref{action}) $\rho_{DM}$ is the DM density for a pressureless fluid which has some non trivial self interaction while here the interaction has been absorbed into the definition of the DM energy. $\rho_{DM}$ in (\ref{action}) is related to the $\{\lambda,\varphi\}$ variables by the relation $\rho_{DM}=\lambda V(\varphi)^2$. Since our knowledge of the DM distribution comes 
through its gravitational effects, definition (\ref{ro}) gives the actual measured DM density.

We also notice that the pressure defined in (\ref{pressure}) acts only along the direction defined by the fluid four velocity, as can be seen from the fact that the four acceleration $a_\mu = u^\nu \nabla_\nu u_\mu$ is identically zero. As a consequence we don't  have anisotropic stresses that in more general situations may be present \cite{Bettoni}.

The effect of the non--minimal coupling is twofold: on the one side it modifies the dark matter energy density while on the other side it introduces a pressure term that would be absent in the standard \lcdm scenario. We expect both these terms to have relevant cosmological consequences at the time and scales of interest, as will be shown below. We anticipate that this model is able, with appropriate potential shapes, to reduce the source of the gravitational potential with the consequence of smoothing out the overdensities at small scales. The equations for a pressureless dust (\ref{llfluid})-(\ref{lffluid}) in the limit of vanishing coupling are independent on the choice for the potential \cite{Iggy} that is to say the density scales like $a^{-3}$ no matter what potential is chosen. This does not represent a problem since once the mechanism that generates the non--minimal coupling is fixed, so is the interaction term which in turns fixes the shape of the potential. We will not investigate the ultimate nature of the non--minimal coupling here but we will give some examples of potentials that could lead to an interesting phenomenology.

\section{Cosmological framework: background and linear perturbations}
\label{sec:cosmology}

We will now study the cosmological consequences of our model. We will assume a flat FRLW universe filled with the dark matter field plus a cosmological constant.\footnote{Since we are interested in cluster/galaxy scales we are not concerned here about the nature of dark energy. We thus make the minimal choice of a cosmological constant.} In terms of the conformal time $d\tau = dt/a$ we have that:
\begin{eqnarray}
\Hc^2 & = & \frac{8 \pi G}{3}\left[\left(\lambda+\epsilon\lint_{,X}\right)V^2(\varphi) -\epsilon\lint + \rho_\Lambda\right],\label{H}\\
\lambda' & = & -a(\tau) V^{-2}(\varphi)\left[V(\varphi)\rho_{,\varphi}+3\Hc(\rho+\epsilon p)\right],\\
\varphi' & = & a(\tau)V(\varphi),
\end{eqnarray}
where primes indicate derivatives with respect to the conformal time and $\rho_\Lambda = 3 \Lambda/(8 \pi G)$.
The equation for $\lambda$ can be rewritten in terms of the more physical quantity $\rho$ defined by equation (\ref{ro}) which is what appears on the right hand side of the Friedman equation \ref{H} with no cosmological constant. In this respect we are defining the density of the dark matter fluid as the quantity which plays the role of gravitational source. With this consideration the previous system of equations is:
\begin{eqnarray}
\Hc^2 & = & \frac{8 \pi G a^2}{3}\left[\rho+ \rho_\Lambda\right]\label{eq:hubble}\\
\rho' & +& 3\Hc(1+\epsilon w(\varphi))\rho=  0\label{eq:continuity}\\
\varphi' & = & a(\tau)V(\varphi)
\label{eq:phi}
\end{eqnarray}
where $w(\varphi)= p/\rho$. The continuity equation can be interpreted as that of a fluid with a field dependent equation of state which in turns means a time dependent equation of state.

We now present the equations for the linear perturbations in the Newtonian gauge for scales which are well inside the horizon following the notation of \cite{Kodama}. For a detailed derivation of the equations we refer to appendix (\ref{app:perturbations}). The system of equations governing the evolution of the linear perturbations is given as usual by the continuity equation, the Euler equation and the Poisson equation:
\begin{eqnarray}
 \delta'(k,\tau)  +  3 \Hc(\tau)\epsilon\left(\frac{\delta p(k,\tau)}{\delta\rho(\tau)}-w(\tau) \right)\delta(k,\tau)+(1+\epsilon w(\tau))kv(k,\tau)&=&0 \label{eq:contpert}\\
 v'(k,\tau) + \Hc(\tau) v(k,\tau) +k\psai(k,\tau)&=&0\label{eq:eulpert}\\
 k^2 \psai(k,\tau) = 4 \pi G a^2 Q(k,\tau)\rho_{mc}(\tau) \delta(k,\tau) ,\label{eq:poiss}
\label{eq:pert}
\end{eqnarray}

\begin{center}
\begin{table}
\label{tab:par}
\centering
\resizebox{1.\columnwidth}{!}{
\begin{tabular}{|c|c|c|c|c|c|}
\hline
  Potential & Pressure & F($\phi$,X) & Effective coupling  & coupling redshift  \\
\hline
  $V(\varphi)=\sqrt{\rho^{DM}_0}$& $p(z)=-(\rho^{DM}_0)^{2}/\ros$  & $1$ & $(\epsilon \rho^{DM}_0/\ros)=1 $& $z_c=5$\\
  $V(\varphi)=\sqrt{\rho^{DM}_0}e^{-\kappa\varphi}$ &$p(z)=- K\, \text{arcsinh}\left[\left(\frac{\Omega_\Lambda/\Omega_{DM}}{(1+z)^3}\right)^{1/2}\right]^4$ & 1 & $(\epsilon \rho^{DM}_0/\ros)=-10^{-3} $&$z_c=5$ \\
  \hline
\end{tabular}}
\caption{Functions and parameters. In the first column are reported the two potentials used in the paper while in the second column are reported the related pressures as a function of the redshift. The third column reports the used value for the scale function F. The value is set to a constant meaning that no scale dependence is present. The last two columns report the value for the effective coupling constant $\epsilon \rho_0^{DM}/\ros$ and the redshift at which the non--minimal coupling is activated.}
\end{table}
\end{center}
where $\delta \equiv \delta \rho/\rho$ and $\rho_{mc}$ is the minimally coupled background DM density. With this formalism the Q function takes the general form $Q(k,\tau)=(1+ \mathcal{O}(\epsilon))$, where the corrections come from the modified background DM density.

From eq. (\ref{eq:contpert}) we can see that the continuity equation is modified in two ways: the last term can lead to a speed up or slow down of the growth of perturbations, depending on the sign of $\epsilon$; the second term on the left hand side is a new genuine effect of this model that closely resembles a dilution term.

The Euler equation is modified in two parts: $\Hc(\tau)$ is modified as in the Friedman equation (\ref{H}) and the gravitational potential is changed as from the Poisson equation. Notice also that despite the presence of an effective pressure, no
 Jeans length appears in the equation. This is a consequence of the time--like character of the pressure term, as noted above. Deviation from the \lcdm model can be parametrized by two functions \cite{AmendolaDE}: $\zeta = (\psai+\fai)/\fai$, that 
characterizes the effects of anisotropic stresses, and $Q$ related to deviation from the standard Poisson equation. The \lcdm  model has $\zeta=0$ and $Q=1$. In our class of models there are no anisotropic stresses and hence $\zeta=0$ as well. However the $Q$ function is in general different from unity and hence it is a measure of the departure from \lcdm on which next generation surveys will cast strong constraints. As said, $Q = 1+\epsilon f(\rho,p)$ in general. Explicit forms for the function $f$ will be given below.

It is a standard procedure to take the derivative of eq.(\ref{eq:contpert}) and using eq.s (\ref{eq:eulpert}) and (\ref{eq:poiss}) to obtain one single equation describing the evolution of the linear perturbations. In our case this gives:
\begin{multline}
\delta''(k,\tau)-3\Hc'(\tau)\epsilon\left(\frac{\delta p(k,\tau)}{\delta \rho(k,\tau)}-w(\tau)\right)\delta(k,\tau)-3\Hc\epsilon\left[\left(\frac{\delta p(k,\tau)}{\delta \rho(k,\tau)}\right)'-w'(\tau)\right]\delta(k,\tau)\\
-3\Hc(\tau)\epsilon\left(\frac{\delta p(k,\tau)}{\delta \rho(k,\tau)}-w(\tau)\right)\delta'(k,\tau) -\epsilon w'(\tau)kv(k,\tau)+(1-\epsilon w(\tau))kv'(k,\tau)=0
\label{eq:deltasecond}
\end{multline}
which reduces to the standard \lcdm equation $\delta''=-kv$ for $\epsilon\rightarrow 0$.
For completeness we give also the Euler equation in real space:
\begin{equation}
v'(x,\tau) + \Hc(\tau)v(x,\tau)=-\nabla \psai(x,\tau) . 
\end{equation}
 Note again that here both $H(\tau)$ and $\psai(x,\tau)$ are modified according to equations (\ref{H}) and (\ref{eq:poiss}) respectively.

\section{Results}
\label{sec:results}

In this section we present the results of the integration of equations (\ref{eq:continuity})-(\ref{eq:deltasecond}) for different choices of the potential which directly translates in different time behaviors for the pressure. In the absence of a clear mechanism that can predict the form of the potential, we consider various examples and derive their possible cosmological implications. We will first explore the case in which the potential for the field $\varphi$ is a constant and then we will consider a decaying exponential potential. These can be seen as
the extrema of the general class of decaying potentials. In fact, all power law potentials would have an intermediate behavior, while a potential linear in the field would give an exponentially decaying time behavior.

As a further work assumption we choose $F=1$. This means that the coupling is only time dependent and not also scale dependent so that as soon as the critical temperature is reached the coupling is active everywhere. This is a crude simplification which enables us to illustrate, in a 
first concrete example, the impact that the dark matter non--minimal coupling illustrated in section (\ref{sec:PDM}) can have on observations. We leave more realistic scale dependent scenarios to future study.
The set of functions and parameters used are reported in table \ref{tab:par}.  We also fix cosmological parameters as follows: $\Omega_\Lambda = 0.76$, $\Omega_{DM} = 0.24$ and $H_0=74$ km/s/Mpc.

\subsection*{Constant potential}

\begin{figure}
\includegraphics
{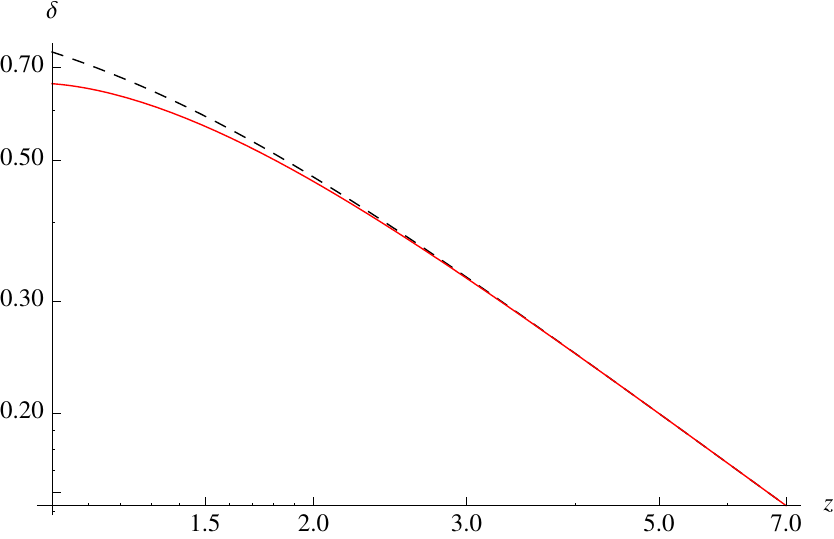}
\includegraphics
{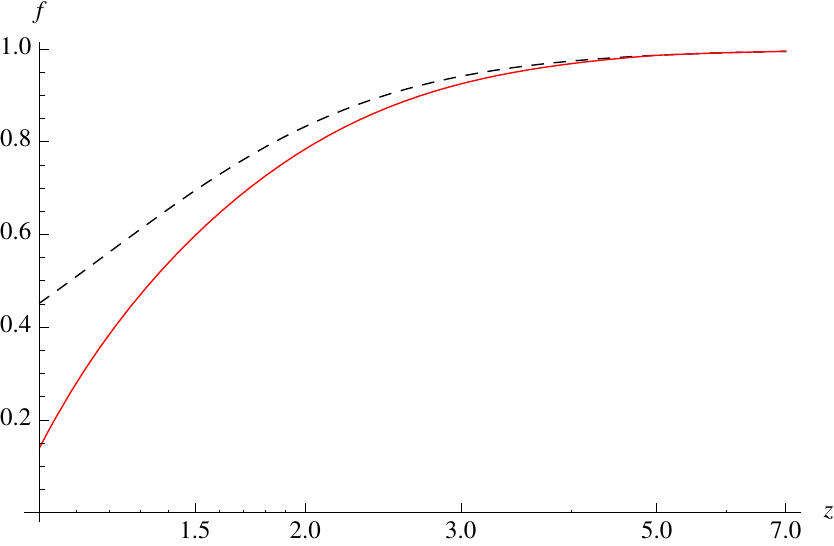}
\caption{(a) Evolution of the density contrast versus redshift in the case of constant potential (left panel, red line) and  (b)  the growth function $f(z)=-d\log\delta/d\log z$ (right panel, red line), compared to the standard \lcdm results (black dashed line), both with $F(X,\varphi)=1$ and $z_c=5$.  The constant $\epsilon_{eff}$ in the case of the constant potential is chosen in order to give a clear idea of the effects of the geometrical interaction term. In particular, in this case, $\epsilon_{eff}=1$.}
\label{fig:delta_cost}
\end{figure}

In the case of a constant potential the effects on the background density evolution are shown 
in figure (\ref{fig:bkg}a) as compared to the evolution of standard cold dark matter. Remarkably the effect of the coupling is to mimic a cosmological constant as can be seen from the background solution:
\begin{eqnarray}
\rho(z)&=&\rho_{dm}^0 (1+z)^3 + \frac{\epsilon_{eff}}{2}\rho_{dm}^0\\
p(z)&=&-\frac{\epsilon_{eff}}{2}\rho_{dm}^0
\end{eqnarray}
where $\epsilon_{eff}=\epsilon \rho_{dm}^0/\ros$ with $\rho_{dm}^0$  the present day density of dark matter, and $\ros$ a reference density characteristic of the scale under analysis.
In this case the background dark matter fluid behaves as if it were composed by two fluids, one standard pressureless fluid plus a fluid with a cosmological constant equation of state $p=-\rho$, as pointed out also in \cite{Iggy} in a different context, so that there is no extra need to include a cosmological term in eq (\ref{H})
The evolution for the density contrast is shown in figure (\ref{fig:delta_cost}a) as compared to that 
 of the density contrast for the \lcdm model. The suppression is enhanced for illustrative purposes but it is clear that the growth of linear perturbation is suppressed with respect to the one in standard \lcdm model, in a way that resembles a \lcdm model with a larger cosmological constant. This is also clear from figure (\ref{fig:delta_cost}b) where growth function $f \equiv -\log \delta/\log z$ for our model decreases faster compared to the \lcdm function.

The $Q$ function, defined in the Poisson equation and responsible for deviation in the gravitational potential has the following form in the case of a constant potential:
\begin{equation}
Q(k,z)= \left(1-\frac{\epsilon_{eff}}{2(1+z)^3}\right) \label{Q}
\end{equation}
which implies that the effective gravitational constant that generates the gravitational potential is reduced at those scales at which the interaction is active.

To conclude, the effect of the gravitational self coupling with a constant dark matter potential is to add an extra contribution analogous to that given by a cosmological constant with the result that the growth of the density contrast is more suppressed than in the standard \lcdm model.

\subsection*{Exponential potential}

\begin{figure}
\includegraphics
{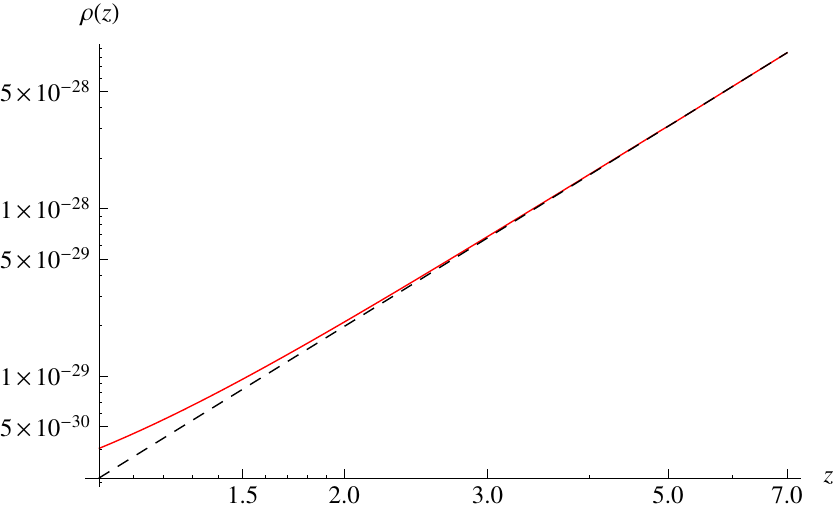}
\includegraphics
{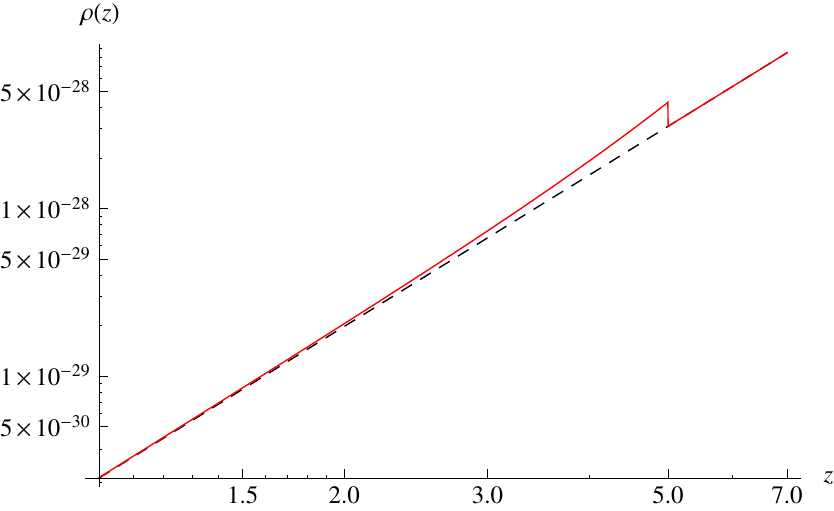}
\caption{Background density plots for (a) the constant potential (left panel, red line) and  (b)  the exponential potential (right panel,red line), compared to standard pressureless dark matter (black dashed line). In both plots we fix $F(X,\varphi)=1$ and $z_c=5$. The density is in $g/cm^3$. The constant $\epsilon_{eff}$ in the case of the constant potential is chosen in order to give a clear idea of the effects of the geometrical interaction term. In particular, in this case, $\epsilon_{eff}=1$.}
\label{fig:bkg}
\end{figure}

The exponential potential acts in a completely different way. The background DM density is shown in figure (\ref{fig:bkg}b) as compared to the evolution of standard  pressureless dark matter. The effect of the non--minimal coupling is to slow down the dilution of the dark matter density as soon as the coupling is switched on.  Notice that the sudden change in the evolution behavior is  a mere consequence of the step function that switches on the coupling only for $t>t_c$.  More realistically, a smoother crossing is expected.
The effect of the coupling fades away with time and the model tends asymptotically to \lcdm. This is a welcome feature as it could possibly boost the number of high redshift clusters observed around $z\sim 2$ without affecting present day halos \cite{clustevol,Lee}, similarly to the scenario pictured in \cite{Baldi,VP} but here relying only on the dark matter non--minimal coupling to gravity. 

In this case the explicit form of the pressure term, as obtained from eq. (\ref{pressure}) is:
\begin{equation}
p(z)=K \text{arcsinh}\left[\left(\frac{\Omega_\Lambda/\Omega_{DM}}{(1+z)^3}\right)^{1/2}\right]^4,
\end{equation}
where
\begin{equation}
K=\frac{1}{2}\left(\frac{3}{2}\right)^4 \rho_0 \left(\frac{3\rho_\Lambda}{8 \pi G}\right)^2.
\end{equation}

\begin{figure}
\includegraphics
{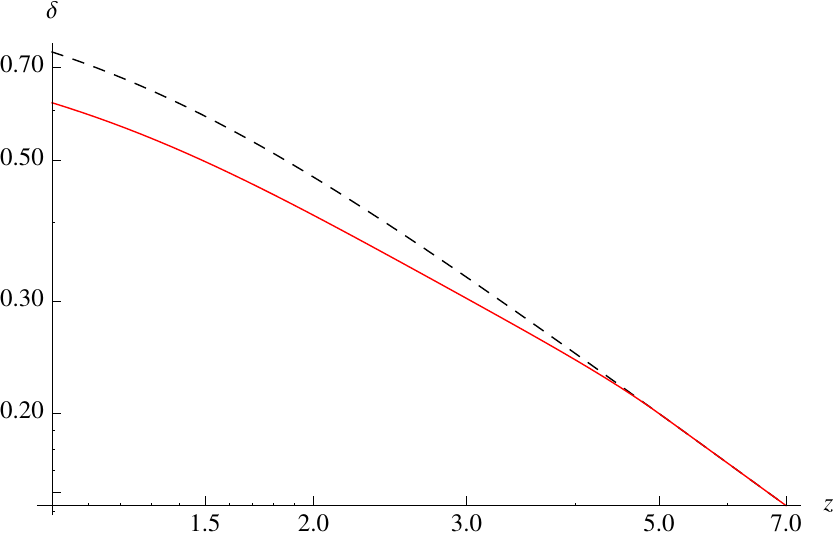}
\includegraphics
{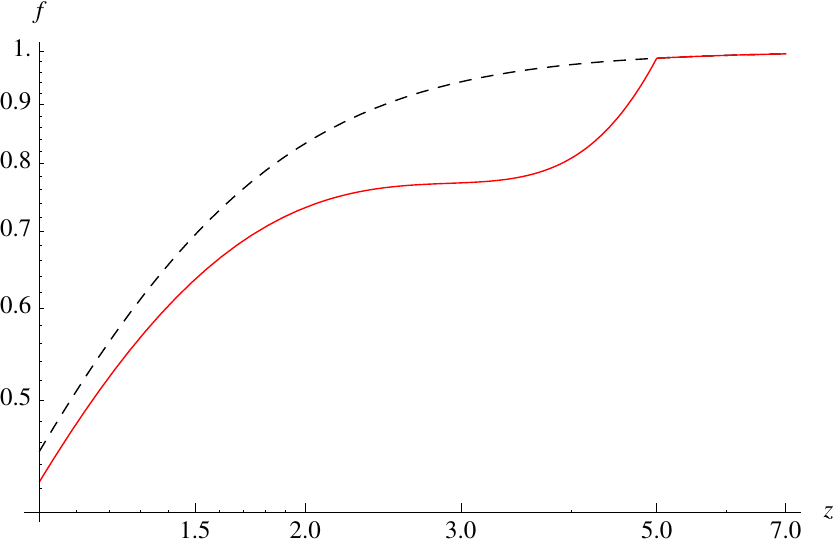}
\caption{(a) Evolution of the density contrast versus redshift in the case of exponential potential, left panel red line and  (b)  the growth function $f(z)=-d\log\delta/d\log z$, right panel red line, compared with the standard \lcdm results (black dashed line), both with $F(X,\varphi)=1$ and $z_c=5$.}
\label{fig:delta_exponential}
\end{figure}

In figure(\ref{fig:delta_exponential}a) the evolution of the density contrast as a function of redshift is plotted . In this case the $Q$ function has a complicated expression, not reported here, due to the non--trivial relation between time and redshift for a \lcdm model. We just comment that also in this case the function is always less than one, thus reducing the gravitational potential.

In figure (\ref{fig:delta_exponential}b) we plot the growth function $f $ as a function of redshift. As in the case of the constant potential, here again we notice that the effect of the coupling is to reduce the growth of linear perturbations. In this case, however, the sign of the coupling constant is opposite. The coupling $\epsilon$ is a phenomenological parameter and different potentials subtend different theories thus the sign of the coupling is not \textit{a priori} determined.  A more mathematical explanation for this fact can be given in the limit of Einstein--de Sitter universe. In this case the equation to be solve for the density can be generally written as
\begin{equation}
\rho'(z) -\frac{3}{z}\left(\rho(z)+\epsilon \rho_0^{DM}z^\alpha\right).
\end{equation}
This equation has the general solution
\begin{equation}
\rho(z)= \rho_0^{DM} z^3 + 3Cz^\alpha
\end{equation}
where $C=-\epsilon/(-3 +\alpha)$ with $\alpha \neq 3$.\footnote{Notice that the case of $\alpha=3$ would simply rescale the coefficient of the homogeneous solution.} For a constant potential $\alpha=0$ and thus the particular solution to the differential equation is positive, hence the density is higher compared to \lcdm. In the exponential case instead $\alpha =6$ and hence the situation is reversed.
 Interestingly, for an exponential potential, this suppression is limited in time: perturbations are maximally suppressed around the time of the switching but asymptotically the model reduces to \lcdm. Again, This can be interesting when trying to get a higher number of halos in the past only, without affecting present abundances.

\section{Conclusions}
\label{sec:conclusions}

Several observational tests coming from future surveys will soon be able to further proof or disproof the $\Lambda$CDM model with respect to non-standard scenarios. In this paper we have addressed the question of whether the dark matter fluid can behave differently at galactic scales rather than at cosmological scales, due to the presence of a non-minimal interaction between dark matter and gravity. We have extended the analysis done in \cite{Bettoni} and we have illustrated for the first time cosmological consequences of such a scenario, both at the background level and within linear perturbation theory. In particular we have shown that a non--minimally coupled dark matter fluid is able to produce two relevant effects:  a pressure term for dark matter able to reduce the growth of structures at small scales, plus an effective interaction term between dark matter and baryons that can explain correlations between the two components of the cosmic fluid. 

In this scenario we have considered the situation in which dark matter, at suitably late times, undergoes some sort of phase transition (\eg analogous to a Bose-Einstein condensation), 
consequently developing a coherence length of a size comparable to that of the local curvature radius, thus becoming non--minimally coupled.

We have studied in details the dark matter pressure term, neglecting the roles of baryons in the present analysis. In particular we have analysed the system for two choices of the dark matter potential that generates the pressure term: a constant potential, resembling a cosmological constant contribution, and an exponential potential. These two choices  are a good sample as all power potentials have intermediate behaviors.

For a constant dark matter potential, the dark matter fluid behaves like the superposition of two fluids, one standard pressureless dust plus a fluid that behaves like a cosmological term, with a consequent suppression of the density contrast at small redshifts. 

In the case of an exponential dark matter potential the effects are mostly relevant near the time of activation of the coupling. The background density is enhanced and the linear growth is suppressed for a limited redshift interval. In fact the pressure decays with redshift so that standard \lcdm evolution is recovered asymptotically.

We also provided the Euler and Poisson equations (\ref{eq:eulpert}), (\ref{eq:poiss}) in a form convenient for N--body simulations for any choice of the potential $V$ together with explicit expressions and predictions for the parameters $Q$ and $\zeta$, 
that characterize the deviations from \lcdm.  In particular we found that $\zeta=0$, meaning that no anisotropic stresses are generated by our model, and $Q < 1$, after the coupling is switched on, thus reducing the gravitational potential. Constraints on these functions has been cast but they are rather weak and strongly model dependent \cite{q1,q2,q3,q4,q5} thus making necessary a direct confrontation between our model and observations in order to cast constraints on the deviations of our model from \lcdm.

For both choices of the potential we have obtained a suppression in the growth of linear perturbations as in the figures (\ref{fig:delta_cost}) and (\ref{fig:delta_exponential}). This is a good indication that this class of models may be a viable possibility to solve some of the \lcdm paradigm problems, like  the core--cusp, the missing satellites, high--z clusters. Of course, a non--linear analysis is required to evaluate these effects.

We have derived the general perturbation equations valid for any $F(\varphi,X)$ and $V(\varphi)$ (see eq.s (\ref{eq:hubble}, \ref{eq:continuity},  \ref{eq:contpert}, \ref{eq:eulpert}, \ref{eq:poiss})) , though we have limited our specific examples to $F(\varphi,X) = 1$, for simplicity. 
Baryons can be included, introducing a much welcome relation between dark matter properties and baryonic features as expected from observations \cite{Salucci}.

We leave to follow-up studies a more realistic and quantitative description, including an environment dependent coupling with a generic $F(\varphi,X)$; observations can be used to put constraints on the free parameters, $\epsilon, \ros$ and on the function $F(\varphi,X)$.

\acknowledgments

The research leading to these results has been funded by the Young SISSA Scientists Research Projects, scheme 2011-2012, promoted by the International School for Advanced Studies (SISSA), Trieste, Italy. DB  and VP wish to thank Ignacy Sawicky for useful discussion. VP is supported by the Marie Curie People IEF. CB acknowledges support by the PD51 INFN Grant.

\appendix

\section{Derivation of the background equations}
\label{app:background}

In this appendix we give full details of the derivation of the equations reported in section (\ref{sec:PDM}). The starting action is
\begin{equation}
S=S_{HE}[g]+S_{DM}[g,\varphi] + \frac{\epsilon}{\ros}\int d^4x\sqrt{-g} X\F \left(X+\frac{1}{2}V^2(\varphi)\right)\label{app:action}.
\end{equation}
The variation with respect to the independent variables $\varphi$, $\lambda$ and the metric $g_{\mu \nu}$ respectively gives:
\begin{multline}
\lambda \square \varphi + \nabla_\mu \lambda \nabla^\mu \varphi - \lambda V(\varphi) V_{\varphi}(\varphi) + \frac{\epsilon}{\ros}\left[\square \varphi \left(\F \left(2X + \frac{1}{2}V(\varphi)^2\right) +F_{\varphi}X\left(X+\frac{1}{2}V^2(\varphi)\right)\right)+\right.\\ 
\left. \nabla_\mu \varphi \nabla^\mu\left(\F\left(2X + \frac{1}{2}V(\varphi)^2\right)+\F_X X\left(X+\frac{1}{2}V^2(\varphi)\right)\right)+\right.\\
\left. \F XV(\varphi)V_\varphi(\varphi)+\F_X X\left(X+\frac{1}{2}V^2(\varphi)\right) \right]=0,
\label{app:KGE}
\end{multline}
\begin{equation}
 X- \frac{1}{2}V(\varphi)^2=0, \label{app:constraint}
\end{equation}
\begin{equation}
G_{\mu \nu} = 8 \pi G T_{\mu \nu} + 8 \pi G T_{\mu \nu}^{Int} \label{app:EFE}
\end{equation}
where
\begin{multline}
T_{\mu \nu}^{Int} = \frac{2\epsilon}{\ros}\left\{ \nabla_\mu \varphi \nabla_\nu \varphi \left[\F\left(2X + \frac{1}{2}V(\varphi)^2\right) \right.\right.+\\
\left.\left.\frac{\partial \F}{\partial X}X\left(X+\frac{1}{2}V^2(\varphi)\right)\right]+ g_{\mu \nu}\F X\left(X+\frac{1}{2}V(\varphi)^2\right)\right\}
\end{multline}
In order to simplify the notation we define the following quantity:
\begin{equation}
 \lint(X,\varphi) \equiv \frac{1}{2}\F T^{\mu \nu}\frac{\nabla_\mu \varphi \nabla_\nu \varphi}{\ros} =\frac{\F}{\ros} X\left(X+\frac{1}{2}V(\varphi)^2\right),
\end{equation}
so that the stress energy tensor can be rewritten as:
\begin{equation}
T_{\mu \nu} = \left(\lambda +\epsilon \lint_X\right)\nabla_\mu \varphi \nabla_\nu \varphi + \epsilon \lint g_{\mu \nu}.
\end{equation}
By a direct confrontation with the shape of the perfect fluid stress energy tensor we can identify the pressure of the field as:
\begin{equation}
p= \lint(X,\varphi).
\end{equation}
Then, using the constraint equation (\ref{app:constraint}) we get:
\begin{eqnarray}
&& \rho = \left(\lambda + \epsilon \lint_X\right)V(\varphi)^2 - \epsilon \lint,\\
&& p=\lint,\\
&& u^\mu = V(\varphi)^{-1}\nabla^\mu \varphi.
\end{eqnarray}
The constraint (\ref{app:constraint}) can be rewritten as:
\begin{equation}
\dot{\varphi} = -V(\varphi), \qquad \dot{}\equiv u_\mu \nabla^\mu  \label{dot}
\end{equation}
while we have that:
\begin{equation}
\vartheta \equiv \nabla_\mu u^\mu = V(\varphi)^{-1}\square \varphi + V_\varphi(\varphi)
\end{equation}
with this set of definitions, equations (\ref{app:KGE})  and (\ref{app:constraint}) can be rewritten as:
\begin{eqnarray}
&& \dot{\lambda} = V^{-2}\left[V(\varphi)\rho_{,\varphi} - \left(\rho +\epsilon p\right)\vartheta\right] \label{eq:lambda},\\
&& \dot{\varphi} = -V(\varphi),
\end{eqnarray}
where we notice that the minus sign in equation (\ref{eq:lambda}) has appeared coming from the definition of derivative (\ref{dot}) and where:
\begin{equation}
\rho_{,\varphi} =
2 \lambda VV_{,\varphi}+\epsilon \left[\lint_{,XX}V^3V_{,\varphi}+ \lint_{,X\varphi}V^2+\lint_{,X}VV_{,\varphi}-\lint_{,\varphi}\right].
\end{equation}
Here we have defined the various terms in the Lagrangian as follows:
\begin{eqnarray}
 \lint & =& \frac{1}{2}\F \frac{V(\varphi)^4}{\ros}, \quad \lint_{,X} = \frac{3}{2}\F\frac{V(\varphi)^2}{\ros}+\frac{1}{2}F_{,X}(X,\varphi)\frac{V(\varphi)^4}{\ros}, \\
 \lint_{,XX}& =& 2\frac{\F}{\ros}+\frac{1}{2}\F_{,XX}\frac{V(\varphi)^4}{\ros}+\frac{1}{2}\F_X\frac{V(\varphi)^2}{\ros}, \\
\nonumber
 \lint_{,X\varphi}& =& \F \frac{V(\varphi)}{\ros}V_{,\varphi}(\varphi)+\frac{1}{2}F_{,\varphi X}\frac{V(\varphi)^4}{\ros}+\\
&+&\frac{1}{2}\F_{,X} V_{,\varphi}(\varphi)\frac{V(\varphi)^3}{\ros}+\frac{3}{2}\F_{,\varphi}\frac{V(\varphi)^2}{\ros},\\
\lint_{,\varphi}& =& \frac{1}{2}\F \frac{V(\varphi)^3}{\ros}V_{,\varphi}(\varphi)+\frac{1}{2}\F_{,\varphi}\frac{V(\varphi)^4}{\ros}.
\end{eqnarray}
Notice that the Klein-Gordon equation for the field $\varphi$ is now an evolution equation for the Lagrangian multiplier $\lambda$ eq (\ref{eq:lambda}). 

\section{Linear cosmological perturbation theory}
\label{app:perturbations}

In this appendix we derive the equations reported in section (\ref{sec:cosmology}) following the notation of Kodama and Sasaki \cite{Kodama}. 
Given a metric $\tilde{g}_{\mu \nu}$, we perturb it around a background solution,
\begin{equation}
\tilde{g}_{\mu \nu} = g_{\mu \nu} + \delta g_{\mu \nu},
\end{equation}
where $\delta g_{\mu \nu} \ll 1$. The stress energy tensor of matter is perturbed in the same way around its perfect fluid form,
\begin{equation}
\tilde{T}^\mu_\nu=T^\mu_\nu + \delta T^\mu_\nu
\end{equation}
To first order in the perturbed quantities the conservation equations $\tilde{\nabla}_\nu\tilde{T}^\nu_\mu=0$ read:
\begin{equation}
\partial_\mu \delta T^\mu_\nu + \Gamma^\mu_{\mu \sigma} \delta T^\sigma_\nu +\delta \Gamma^\mu_{\mu \nu} T^\sigma_\nu - \Gamma^\sigma_{\mu \nu}\delta T^\mu_\sigma - \delta \Gamma^\sigma_{\mu \nu} T^\mu_\sigma =0 .
\end{equation}
In the Newtonian gauge the previous equations give for the time component
\begin{equation}
\delta \rho' (\tau,{\bf k})+ h\left(v(\tau,{\bf k})k+3\fai'(\tau,{\bf k})\right) + 3 \Hc(\tau) \delta \rho(\tau,{\bf k}) + 3\epsilon \Hc(\tau) \delta p(\tau,{\bf k})=0,
\end{equation}
and 
\begin{equation}
 h(\tau) v'(\tau,{\bf k}) + \left(4\Hc(\tau) h(\tau) + h'(\tau)\right)v(\tau,{\bf k}) - \epsilon k\delta p(\tau,{\bf k}) - h(\tau)k\psai(\tau,{\bf k}) + \frac{2}{3}\pi_T(\tau,{\bf k}) k=0,
 \end{equation}
for the spatial component. Perturbing the Einstein field equations (\ref{app:EFE}) one obtains a set of coupled equations that relates the perturbed metric quantities to the perturbed matter sources. The $(0,0)$ component describes how the perturbed matter density sources the gravitational potentials
\begin{equation}
 3\Hc(\tau)^2\psai(\tau,{\bf k}) - 3\Hc(\tau)\fai'(\tau,{\bf k}) -k^2\fai(\tau,{\bf k}) = -4 \pi G a^2\delta \rho(\tau,{\bf k}),
\end{equation}
the $(0,i)$ component relates gravitational potentials to the peculiar velocity of the matter,
\begin{equation}
 k\Hc\psai(\tau,{\bf k}) -k\fai'(\tau,{\bf k})= 4 \pi G a^2 h(\tau)v(\tau,{\bf k}),
\end{equation}
while the trace of the $(i,j)$ component relates gravitational potentials to the isotropic pressure,
\begin{multline}
\left(2a'' - \Hc(\tau)^2 \right)\psai(\tau,{\bf k}) + \Hc(\tau) \psai'(\tau,{\bf k}) - \frac{1}{3}k^2\psai(\tau,{\bf k})-\frac{1}{3}k^2\fai(\tau,{\bf k}) +\\
-2\Hc(\tau) \fai'(\tau,{\bf k}) -\fai''(\tau,{\bf k})= 4 \pi G  a^2\epsilon \delta p(\tau,{\bf k}).
\end{multline}
 As usual, the sum of the gravitational potentials is related to the anisotropic stresses: 
\begin{equation}
 -k^2\left(\psai(\tau,{\bf k}) +\fai(\tau,{\bf k})\right)=8 \pi G a^2\epsilon p(\tau) \pi_T(\tau,{\bf k}) 
\end{equation}
where $\psai$ is the Newtonian potential. Since the stress energy tensor is diagonal and the perturbed equations do not contain terms that mix different components, then $\pi_T=0$ and hence $\fai = -\psai$. 

Following \cite{Iggy} we can add another equation to the system, perturbing the constraint equation (\ref{app:constraint}). This is not independent as it turns out to be the equation for the velocity potential $v = V(\varphi)^{-1}k/a \delta \varphi$. The constraint equation is
\begin{equation}
\tilde{\varphi}' = \tilde{g}^{00}\tilde{u}_0\partial_0 \tilde{\varphi}\,,
\end{equation}
where the tilde indicates the physical quantity containing also the perturbations. When perturbed in Newtonian gauge this gives:
\begin{equation}
\delta \varphi' = a\left[V_\varphi(\varphi) \delta \varphi + \psai V(\varphi)\right] .
\end{equation}
Hence the system of equations to be solved is the following:
\begin{eqnarray}
&& \delta \varphi' =a\left( V_\varphi(\varphi) \delta \varphi + \psai V(\varphi)\right)\label{B6}\\
&& \delta \rho' +h\left(kv - 3\psai'\right)+3\Hc \left(\delta \rho + \epsilon \delta p\right)=0\\
&&v' + \left(\Hc +\frac{\epsilon}{2}\frac{p'}{\rho}\right)v -k\psai = \epsilon k\frac{\delta p}{\rho}\label{B8}\\
&& k^2\psai + 3\Hc \left(\psai' + \Hc\psai \right)=4 \pi G a^2\delta \rho\label{B9}\\
&& k\left(\psai' + \Hc \psai\right) = 4 \pi G a^2 h v\\
&& \psai''+  \Hc \psai' + \left(2\Hc'+\Hc^2\right)\psai = 4 \pi G a^2 \epsilon\delta p .\label{B11}
\end{eqnarray}
This  set of equations is formally equivalent to the standard one apart from the first equation, derived from the perturbed constraint given by eq. (\ref{app:constraint}). However in the class of models under consideration extra relations exist that link together some of the variables. We have:
\begin{eqnarray}
\nonumber
&& v = V(\varphi)^{-1}k\delta \varphi,\\
&& \delta p = p_\varphi \delta \varphi .  
\label{relations}
\end{eqnarray}
Notice that the system (\ref{B6})-(\ref{B11}) is redundant as we have 6 equations and 5 unknown variables. In what follows we rewrite Poisson and Euler equations (\ref{B8}),(\ref{B9}), in such a way that the modifications due to the non--minimal coupling are manifest.

We are interested in the dynamics at scales well inside the horizon, $\lambda \equiv \Hc/k \ll 1$ and such that the characteristic evolution time for the potential $\psai$ is of the order of $\Hc^{-1}$.
The system of equations governing the evolution of the perturbations in this limit reads:
\begin{eqnarray}
&& \delta \varphi' =a\left( V_\varphi(\varphi) \delta \varphi + \psai V(\varphi)\right)\\
&&\delta \rho' +hkv+3\Hc \left(\delta \rho + \epsilon p_\varphi \delta \varphi\right)=0\\
&&v' + \left(\Hc +\epsilon\frac{p'}{\rho}\right)v -k\psai = \epsilon k\frac{\delta p}{\rho}\label{app:Euler}\\
&&  k^2\psai =4 \pi G a^2\delta \rho \label{app:Poisson}\\
&& \psai'+\Hc \psai = 4 \pi G a^2 h v/k\\
&&\psai'' + \Hc\psai' + (2\Hc'+\Hc^2)\psai = 4 \pi G a^2 \epsilon \delta p .
\label{se:subh}
\end{eqnarray}

Let's consider now Euler equation (\ref{app:Euler}). Using (\ref{relations}) the two terms proportional to $\epsilon$ cancel giving formally the same expression as in the standard case:
\begin{equation}
v'(k,\tau)+ \Hc(\tau) v(k,\tau) -k\psai(k,\tau) = 0.
\label{eq_app:euler}
\end{equation}
However here $\Hc$ is the modified Hubble expansion rate  obtained by solving equation (\ref{H}). Eq. (\ref{eq_app:euler}) is therefore actually non standard both in the friction term and in the gravitational potential that feeds it, which, from eq. (\ref{app:Poisson}), can be rewritten as: 
\begin{equation}
 k^2 \psai(k,\tau) = 4 \pi G a^2 Q(k,\tau)\rho(\tau)\delta(k,\tau),
\label{eq_app:Poisson}
\end{equation}
where the $Q$ function measures deviation form the \lcdm model which has $Q=1$ and where $\delta$ is the dimensionless density contrast $\delta \equiv \delta \rho/\rho$. Explicit shapes for this function are given in section (\ref{sec:results}).

The continuity equation is usually rewritten in terms of the dimensionless density contrast $\delta$. This gives:
\begin{equation}
\delta'(k,\tau) + 3\epsilon\Hc(\tau)\left(\frac{\delta p(k,\tau)}{\delta\rho(k,\tau)}-w(\tau)\right)\delta(k,\tau) + \left(1+\epsilon w(\tau)\right)kv(k,\tau)=0
\end{equation}
where $w=p/\rho$. The full analysis of these equations is carried out in section (\ref{sec:PDM}) and  (\ref{sec:cosmology}).

\end{document}